\documentclass[aps,pra,twocolumn,superscriptaddress,showpacs]{revtex4-1}
\usepackage{graphicx}
\usepackage{amsmath}
\usepackage{epsfig}

\newcommand{\beq}{\begin{eqnarray}}
\newcommand{\eeq}{\end{eqnarray}}

\begin{document}

\title{A temporal steering inequality}
\author{Yueh-Nan Chen}
\email{yuehnan@mail.ncku.edu.tw}
\affiliation{Department of Physics and National Center for Theoretical Sciences, National
Cheng-Kung University, Tainan 701, Taiwan} \affiliation{CEMS, RIKEN,
Wako-shi, Saitama 351-0198, Japan}
\author{Che-Ming Li}
\email{cheming.li@gmail.com}
\affiliation{Department of Engineering Science, National Cheng-Kung University, Tainan
City 701, Taiwan}
\author{Neill Lambert}
\email{nwlambert@riken.jp}
\affiliation{CEMS, RIKEN, Wako-shi, Saitama 351-0198, Japan}
\author{Shin-Liang Chen}
\affiliation{Department of Physics and National Center for Theoretical Sciences, National
Cheng-Kung University, Tainan 701, Taiwan}
\author{Yukihiro Ota}
\affiliation{CEMS, RIKEN, Wako-shi, Saitama 351-0198, Japan}
\author{Guang-Yin Chen}
\affiliation{Department of Physics, National Chung Hsing University,
Taichung 402, Taiwan}
\author{Franco Nori}
\affiliation{CEMS, RIKEN, Wako-shi, Saitama 351-0198, Japan}
\affiliation{Physics Department, University of Michigan, Ann Arbor, MI
48109-1040, USA}

\date{\today }

\begin{abstract}
Quantum steering is the ability to remotely prepare different quantum states
by using entangled pairs as a resource. Very recently, the concept of
steering has been quantified with the use of inequalities, leading to
substantial applications in quantum information and communication science.
Here, we highlight that there exists a natural temporal analogue of the
steering inequality when considering measurements on a single object at
different times. We give non-trivial operational meaning to violations of
this temporal inequality by showing that it is connected to the security
bound in the BB84 protocol and thus may have applications in quantum
communication.
\end{abstract}

\maketitle

\section{Introduction}

Non-locality is one of the most striking concepts of quantum mechanics in
that it defies our intuition about space and time. Its history can be traced
back to such early works as those of Einstein, Podolsky, and Rosen \cite{EPR}%
. After Bell proposed his famous test of non-locality in 1964 \cite{Bell},
almost all experimental implementations have yielded violations of his
inequalities \cite{Aspect}. Motivated by the advance of quantum information
science, quantum non-locality as a resource has been further studied in the
form of entanglement and entanglement measures. Recently, an inequality to
delineate quantum steering from other non-local properties was proposed \cite%
{Wiseman,Caval, Smith} and tested in a range of experiments \cite{Smith,
experiments}. Steerability has since then been further investigated with
all-versus-nothing measures \cite{JingLing}, and has been utilized as a way
to characterize and visualize the state-space of two-qubit systems~\cite%
{Jevtic}. In combination these different concepts (Bell non-locality,
steerability, and entanglement) form a hierarchy and enable one to
categorize different non-local properties of quantum states.

Moving away from the notion of non-locality, Leggett and Garg in 1985
derived an inequality~\cite{LG} to test the assumption of ``macroscopic
realism'' on a single object. An experimental violation of this inequality
has been observed in a large range of systems over the last few years~\cite%
{SC,EmaryReview}. In addition, the Leggett-Garg (LG) inequality can also be
applied to microscopic systems \cite{Lambert1, Lambert2, SR1, SR2}, as a
tool to examine the quantum coherent dynamics therein. For two-level
systems, there is a one-to-one correspondence between the LG inequality and
the Bell-CHSH inequality \cite{Brukner}, a fundamental consequence of the
Choi-Jamiolkowski isomorphism\cite{choi}.

Since the steering inequality is fundamentally linked to the notion of
non-locality we then ask the following question: Does there exist a temporal
scenario or analogue of the steering inequality, as implied by the
Choi-Jamiolkowski isomorphism, and does it have any non-trivial
implications?  We start by showing, in Section II, that there does exist such an
analogue, and in Section III give some simple examples of its behavior.  To give a non-trivial operational meaning to this temporal
steering inequality in Section IV we show that, for a noisy channel, the upper bound on
the noise which limits the observation of a violation of temporal steering
exactly corresponds to the optimal upper limit on the allowable noise in the
BB84 quantum cryptography scheme \cite{BB84}.  In Section V we discuss how spatial and temporal steering
can be distinguished.

\begin{figure}[th]
\includegraphics[width=8cm]{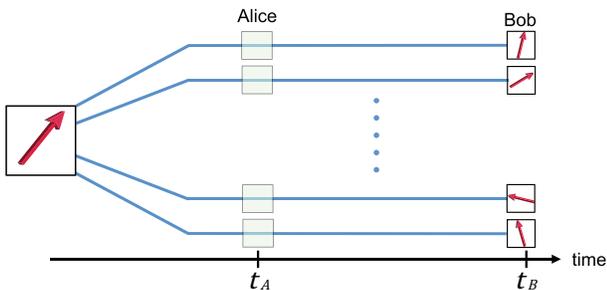}
\caption{(Color online) Temporal scenario of the steering inequality. An
object may be sent into different channels with probability distribution $q_{%
\protect\lambda }$. Alice claims that the non-invasive measurement is
performed at the earlier time $t_{A}$, whereas Bob performs the trusted
quantum measurement at the later time $t_{B}$.}
\label{fig1}
\end{figure}

\section{Formulation}

First, consider a quantum channel through which a system is sent to Bob. At
an intermediate point of the channel, Alice can perform some operations,
including measurements on the system, before Bob receives the system and
performs his measurement. The state of the system is characterized by a set
of observables. In this setting, Alice measures the observable $A_{i}$ at
time $t_{\mathrm{A}}$, and subsequently Bob measures the observable $B_{j}$
at $t_{\mathrm{B}}$. The subscripts $i$ and $j$ are the particular choice of
observable each makes. For example, for a two-level system, when Bob
performs his measurements with a mutually-unbiased basis (e.g., Ref.~%
[\onlinecite{Wootters;Fields:1989}]), the corresponding observables are $%
B_{1}=\sigma _{z}$, $B_{2}=\sigma _{x}$, and $B_{3}=\sigma _{y}$. When the
measurement results of $A_{i}$ and $B_{j}$ are, respectively, $a$ and $b$,
the joint probability distribution of this result is
\begin{eqnarray}
&&P^{\mathrm{Q}}(A_{i,t_{\mathrm{A}}}=a,\,B_{j,t_{\mathrm{B}}}=b)  \notag \\
&&\quad \quad =P^{\mathrm{Q}}(A_{i,t_{\mathrm{A}}}=a)P^{\mathrm{Q}}(B_{j,t_{%
\mathrm{B}}}=b|A_{i,t_{\mathrm{A}}}=a).  \label{eq:q_joint_prob}
\end{eqnarray}%
We explicitly write the measurement times as subscripts of the observables.
The conditional probability $P^{\mathrm{Q}}(B_{j,t_{\mathrm{B}}}=b|A_{i,t_{%
\mathrm{A}}}=a)$ is an expectation value of a projector (or a positive
operator related to a positive operator-valued measure) with respect to a
density matrix depending on a Alice's measurement result. In other words,
this quantity contains the backaction from Alice's measurement. This
backaction can change the quantum dynamics after the measurement.

Now, to obtain a temporal steering-inequality, we consider an alternative
model to describe the aforementioned scenario. In this model, Bob receives a
system that is sent to him through one of several different channels, chosen
randomly, as seen in Fig.~\ref{fig1}. Furthermore, the choice of Alice's
measurement observable has no influence on the state of the system Bob
receives, apart from furnishing information about which channel it is sent
through. In the temporal scenario this can only arise from variations of
three possible scenarios; first, Alice simply does not have access to Bob's
system, and is ``making-up'' her measurement results. Second, Alice does
have access to Bob's system but the influence of her measurement choice is
washed out by the noise in the channel, before Bob receives the system.
Third, Alice cannot measure Bob's qubit, but can determine something about
which channel the system passed through.

The above setting can be regarded as a multi-channel protocol with
probability distribution $q_{\lambda }$, as seen in Fig.~\ref{fig1}. The
classical random variable $\lambda $ specifies a given type of channel. Both
Alice and Bob do not have a priori knowledge about $q_{\lambda }$. We assume
that: (i) Bob trusts only his measurement results, and (ii) Alice's choice
of measurement has no influence on the state Bob receives. Instead of Eq.~(%
\ref{eq:q_joint_prob}), the joint probability distribution can be written as
\begin{eqnarray}
P(A_{i,t_{A}} &=&a,B_{j,t_{B}}=b)  \notag \\
&=&\sum_{\lambda }q_{\lambda }P_{\lambda }(A_{i,t_{\mathrm{A}}}=a)P_{\rho
_{\lambda }}^{\mathrm{Q}}(B_{j,t_{\mathrm{B}}}=b)\text{,}
\label{eq:ts_joint_prob}
\end{eqnarray}%
where $\sum_{\lambda }q_{\lambda }=1$. The conditional probability
\mbox{ \( P_{\lambda}(A_{i,t_{\rm A}}=a)
\)} is associated with Alice's (non-invasive) measurement at time $t_{%
\mathrm{A}}$. The quantum state for the channel $\lambda $ is evolved into $%
\rho _{\lambda }$ at time $t_{\mathrm{B}}$. Bob obtains the conditional
probability \mbox{
\(
P^{\rm Q}_{\rho_{\lambda}}(B_{j,t_{\rm B}}=b)
\)} as quantum mechanics gives (i.e., an expectation value with respect to $%
\rho _{\lambda }$).

Using Eq.~(\ref{eq:ts_joint_prob}) we can derive the temporal analogue of
the steering inequality. Hereafter, we focus on the case when the object is
a two-level (or a two-valued) system in which the observable takes either $%
+1 $ or $-1$. The observables $A$ and $B$ are the Pauli matrices. We stress
that the formula (\ref{eq:ts_joint_prob}) is essentially the same as in the
context of the (spatial) quantum steering inequality\thinspace \cite{Wiseman}%
, where the notion of non-invasive measurement is the analogue to locality.
Following the techniques in Ref.~\onlinecite{Smith}, the temporal steering
inequality is
\begin{equation}
S_{N}\equiv \sum_{i=1}^{N}E\left[ \left\langle B_{i,t_{\mathrm{B}%
}}\right\rangle _{A_{i,t_{\mathrm{A}}}}^{2}\right] \leq 1,
\label{eq:ts_ineq}
\end{equation}%
where $N$($=2$ or $3$) is the number of mutually-unbiased measurements that
Bob implements on his qubit, and
\begin{equation}
E\left[ \left\langle B_{i,t_{\mathrm{B}}}\right\rangle _{A_{i,t_{\mathrm{A}%
}}}^{2}\right] \equiv \sum_{a=\pm 1}P(A_{i}=a)\left\langle B_{i,t_{\mathrm{B}%
}}\right\rangle _{A_{i,t_{\mathrm{A}}}=a}^{2},
\end{equation}%
with
\beq P(A_{i,t_{\mathrm{A}}}=a)\equiv \sum_{\lambda }q_{\lambda }P_{\lambda
}(A_{i,t_{\mathrm{A}}}=a),
\eeq
and Bob's expectation value conditioned on
Alice's result is defined as
\begin{equation}
\left\langle B_{i,t_{\mathrm{B}}}\right\rangle _{A_{i,t_{\mathrm{A}%
}}=a}\equiv \sum_{b=\pm 1}b\,P(B_{i,t_{\mathrm{B}}}=b\mid A_{i,t_{\mathrm{A}%
}}=a).
\end{equation}%
The inequality (\ref{eq:ts_ineq}) comes from the fact two observables in a
mutually-unbiased basis are non-commutative. If the inequality is violated,
it implies Alice's choice of measurement basis influences Bob's measurement
outcomes, and that the channel has not erased the influence of this choice.

\section{Examples}

For illustrative purposes, we first consider an example of a single qubit
that undergoes Rabi-oscillations due to the dynamics given by the
Hamiltonian evolution $H^{\prime }=\hbar g(\sigma _{+}+\sigma _{-})$, where $%
g$ is the Rabi frequency, and $\sigma _{+}$ and $\sigma _{-}$ are the
raising and lowering operators of the qubit. In addition, the qubit is also
subject to an intrinsic Markovian decay process in Lindblad form, so that
the total evolution of the density matrix can be expressed as
\begin{equation}
\dot{\rho}=\frac{1}{i\hbar }[H^{\prime },\rho ]+\frac{\gamma }{2}(2\sigma
_{-}\rho \sigma _{+}-\sigma _{+}\sigma _{-}\rho -\rho \sigma _{+}\sigma
_{-}),
\end{equation}%
where $\gamma $ is the decay rate of the qubit. Assuming that the qubit is
initially in the mixed state,
\begin{equation}
\rho _{t=0}=\left(
\begin{array}{cc}
\frac{1}{2} & 0 \\
0 & \frac{1}{2}%
\end{array}%
\right) ,
\end{equation}%
one first performs a 
measurement along the $\widehat{x}$, $\widehat{y}$, or $\widehat{z}$
direction at time $t_{A}=0$, to mimic the action of Alice in Fig.~\ref{fig1}%
. After the system evolves to time $t$, the second measurement is
implemented along a mutually unbiased basis. The various measurement
outcomes are sorted and arranged conditionally, and the steering parameter
is calculated.
\begin{figure}[th]
\includegraphics[width=8cm]{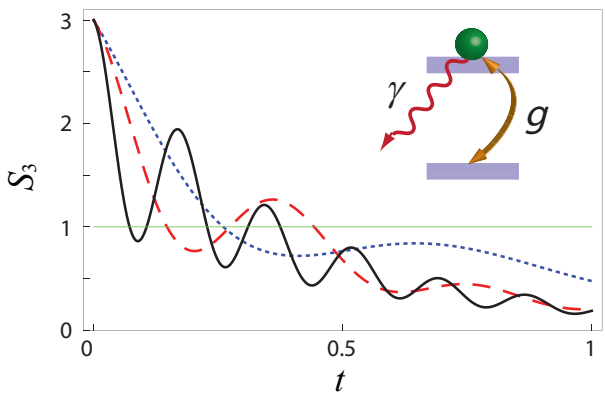}
\caption{(Color online) The temporal steering parameter $S_{3}$ of a qubit
initially prepared in the mixed state undergoes Rabi-oscillations with the
coherent Rabi frequency $g$ and the intrinsic decay rate $\protect\gamma $.
In all figures $\hbar =1$. The black-solid, red-dashed, and blue-dotted
curves represent the results of $g=9\protect\gamma ,$ $4\protect\gamma $,
and 2$\protect\gamma $, respectively.}
\label{fig2}
\end{figure}

In Fig.~\ref{fig2}, we plot the steering parameter $S_{3}$ as a function of
the evolution time. For short times ($t\rightarrow 0$), the steering
parameter always violates the bound and reaches the value of $3$. To
understand this, let us recall the spatial steering inequality. Suppose
Alice and Bob share a Werner state of visibility $V$, \beq \rho _{w}=V\left|
\psi _{-}\right\rangle \left\langle \psi _{-}\right| +(1-V)/4,\eeq where $%
\left| \psi _{-}\right\rangle \left\langle \psi _{-}\right| $ is the Bell
singlet state. After Alice performs her measurements (along the $\widehat{x}$%
, $\widehat{y}$, and $\widehat{z}$ directions), some of the possible
reduced, measurement-conditioned density matrices of Bob's system before
Bob's measurements are made can be written as follows:
\begin{eqnarray}
\rho _{\sigma_z, +1}&=&\left(
\begin{array}{cc}
\frac{1-V}{2} & 0 \\
0 & \frac{1+V}{2}%
\end{array}%
\right), \rho _{\sigma_x, +1}=\left(
\begin{array}{cc}
\frac{1}{2} & -\frac{V}{2} \\
-\frac{V}{2} & \frac{1}{2}%
\end{array}%
\right).
\end{eqnarray}%
Since the steering parameter is the summation of the results along different
unbiased bases, it is the differences in the coherence terms ($V$ in the
off-diagonal element) that give the violation of the inequality.

Returning to the temporal analogue of the steering inequality, similar
results occur. When Alice performs the first measurements along a certain
basis, the qubit is projected into the corresponding state. From the
viewpoint of a\textit{\ }given basis (for example, the $\widehat{z}$ basis),
the measurement creates coherence in the density matrix if the measurement
is along another mutually unbiased $\widehat{x}$ or $\widehat{y}$ direction.
If Bob immediately performs the second measurements after Alice's
measurement, the influence of Alice's choice of measurement is as large as
it can be, resulting in the violations of the bound. Another phenomenon
worth mentioning is that the violation may re-occur at later times if the
frequency $g$ is strong enough (compared with the decay rate $\gamma $).
This feature resembles a similar effect in the Leggett-Garg inequality,
where violations are periodic in time. In Appendix A, we consider an
extension of this example, where revivals also occur due to strong
interactions with a quantum environment. Ultimately, a violation
of the temporal steering inequality means that there are significant quantum
correlations between measurements at different times. However, unlike in the
Leggett-Garg and Bell inequality cases one should be wary of implementing
any version of the steering inequality as a kind of quantum witness, as it
is of course trivially violated by a two-partite classical hidden variable
model. A summary of the various spatial and temporal inequalities are given
in Table \ref{TAB:overview}.

\begin{table*}[tb, floatfix]
\begin{center}
\begin{tabular}{|c|c|}
\hline
CHSH Inequality & Leggett-Garg Inequality \\
~ & ~ \\
~$|\langle B_1 A_1 \rangle +\langle B_1A_2 \rangle+ \langle B_2A_1
\rangle-\langle B_2A_2 \rangle|\leq 2$~ & ~$|\langle A_{i,t_2}A_{i,t_1}
\rangle+\langle A_{i,t_3} A_{i,t_2} \rangle+\langle A_{i,t_4} A_{i,t_3}
\rangle-\langle A_{i,t_4} A_{i,t_1} \rangle|\leq 2 $~ \\
~ & ~ \\ \hline\hline
Steering inequality & Temporal steering inequality \\
~ & ~ \\
$\sum_{i=1}^{N}E\left[ \left\langle B_{i}\right\rangle _{A_{i}}^{2}\right]
\leq 1 $ & $\sum_{i=1}^{N}E\left[ \left\langle B_{i,t_{\mathrm{B}
}}\right\rangle _{A_{i,t_{\mathrm{A}}}}^{2}\right] \leq 1$ \\
~ & ~ \\ \hline
\end{tabular}%
\end{center}
\caption{ A summary of the definitions of the various spatial and temporal inequalities. Particular examples of the spatial inequalities (the Bell inequality and the steering inequality) are shown on the left, with their corresponding temporal analogues on the right.}
\label{TAB:overview}
\end{table*}


\section{Quantum Cryptography}

It is well known that the spatial steering inequality has some promising
applications in quantum communication and quantum cryptography \cite%
{app,Smith}. Here, we discuss how the temporal steering inequality has
directly analogous applications. In particular, as with the spatial steering
inequality \cite{app} and the CHSH inequality\cite{cryptography}, the
temporal steering inequality can be used to directly test the suitability of
a quantum channel for certain quantum cryptography protocols. However,
unlike the spatial steering inequality, one does not need to resort to
entanglement based schemes but can directly work with the BB84 and related
protocols \cite{BB84,app,1sided}.

Typically, in the BB84 protocol one needs to check whether a state sent from
Alice to Bob, from which they wish to construct their private key, is being
measured by a third person, Eve. To do this, Alice and Bob have to compare
their measurement results using a sub-ensemble of their qubits. The
eavesdropping by Eve is equivalent to an environment acting on the quantum
state, or losses in a noisy channel. Thus in any real implementation of BB84
there is an upper-limit to how noisy the channel can be; otherwise the
possible effect of Eve's measurements cannot be distinguished from that
noise.

As an example, let us evaluate the average bit error rate that Alice and Bob
will find when the channel is influenced by Eve's measurements. Alice
performs her projective measurement on an initial state $\rho _{0}$,
producing the state $\rho _{\mathrm{A},\mu }$, with $\mu =(i,a)$. The
subscript $\mu $ represents the choice of measurement direction ($i=\widehat{%
z},\,\widehat{x}$) and the measured value ($a=\pm 1$). Alice then sends this
state to Bob through a quantum channel, during which Eve tries to eavesdrop
and measure the state of the system. We denote the state Bob receives as $%
\rho _{\mathrm{B},\mu }$. The average bit error rate can be written by
\begin{equation}
R_{\mathrm{err}}=\frac{1}{2}\sum_{\mu }P^{\mathrm{Q}}(A_{i,t_{\mathrm{A}}}=a)%
\left[ 1-\mathrm{Tr}(\sqrt{\rho _{\mathrm{A},\mu }}\rho _{\mathrm{B},\mu }%
\sqrt{\rho _{\mathrm{A},\mu }})\right] .
\end{equation}%
During the eavesdropping, Eve randomly measures the qubit along the $%
\widehat{z}$ ($\widehat{x}$) direction with probability $q$ ($p$) or does
nothing with probability $1-(p+q)$ (under the constraint of $p+q\leq 1$).
The effect of this process on the state that Bob receives can be shown to be
(see Appendix C),
\begin{equation}
R_{\mathrm{err}}=\frac{1}{4}(p+q).  \label{eq:error_BB84}
\end{equation}

What is the minimum error that Eve can introduce and still gain sufficient
information to capture the shared key of Alice and Bob? The optimal upper
bound~\cite{cryptography} allowable in a quantum channel so that the channel
is still useful for BB84, and in corollary the minimum error rate Eve can
induce while still extracting useful information, is in the case when the
effect of Eve's actions in each basis is equal, i.e., when $p=q$. When Eve
adopts a strategy that relies just on independent attacks on each qubit the
optimal scenario was found~\cite{cryptography,cryptographyOLD} to be set by
the equation $2(1-2R_{\mathrm{err}})^{2}-1=0$, which gives a threshold error
rate of $R_{\mathrm{err}}(p=q)=0.146447$.

\begin{figure}[th]
\includegraphics[width=7.5cm]{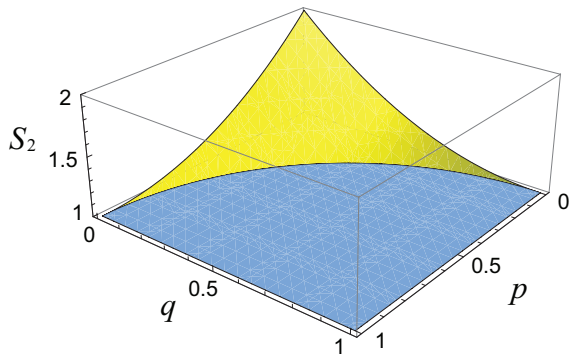}
\caption{(Color online) The steering parameter $S_{2}$ under the disturbance
by the third party, Eve, with the probabilities of measurements $q$ (along
the $\widehat{z}$ direction) and $p$ (along the $\widehat{x}$ direction).}
\label{fig4}
\end{figure}

How are BB84 and the error rate of a channel related to temporal steering?
Since Eve's action can be described as a quantum channel (see appendix B) we can
see a direct relationship between the error rate and the violation of the
temporal steering inequality; if the loss in the channel is too large the
steering inequality is not violated. To make things concrete we focus on the
temporal steering parameter for $N=2$, This is a reasonable choice, because
in the standard BB84 scheme one performs measurements along either $\widehat{%
z}$ or $\widehat{x}$ directions. One finds that, for the quantum channel
described above,
\begin{equation}
S_{2}=(1-p)^{2}+(1-q)^{2}.  \label{eq:S2_BB84}
\end{equation}%
As shown in Fig.~\ref{fig4}, Eq.~(\ref{eq:ts_ineq}) implies an upper bound
of $S_{2}$ equal to unity.

Considering the scenario mentioned above, when Eve's measurements are
equally distributed ($p=q$), from the roots of setting Eq. [\ref{eq:S2_BB84}%
] equal to unity, one also finds that the steering inequality reduces to $%
2(1-2R_{\mathrm{err}})^2-1=0$: The steering inequality bound and the BB84
threshold are equivalent. In other words, the boundary of steerability is
set by the optimal error rate for the BB84 protocol, and \emph{vice versa}.
A similar result was also observed by extending BB84 into an entanglement
utilizing protocol, and calculating the violation of a Bell inequality\cite%
{Ekert}.

However, $R_{\mathrm{err}}(p=q)=0.146447$ is still not the minimum noise
that Eve can induce. She can resort to so-called ``coherent'' attacks where
she can access multiple qubits and operate on them at the same time. In this
case \cite{cryptography} it was found that the minimum error rate was $11\%$%
. Recently Branciard \emph{et al. }\cite{app} showed that the key length in
that case could be mapped to a different type of spatial steering inequality
based on conditional entropies between Alice and Bob's qubits, in analogy to
the entropic Bell and Leggett-Garg inequalities \cite{entropicBell,
entropicLI, entropic}. One can again map this entropic steering inequality
into the temporal domain, which has the same error bound as the spatial one,
producing a temporal entropic steering inequality:
\begin{equation*}
S_{2}^{(E)}\equiv \sum_{i=1}^{2}H(B_{i,t_{B}}|A_{i,t_{A}})\geq 1,
\end{equation*}%
where the average conditional entropies are defined \beq
H(B_{i,t_{B}}|A_{i,t_{A}})=\sum_{a=\pm 1}P(A_{i}=a)H(B_{i}|a),\eeq and \beq
H(B_{i}|a)=-\sum_{b=\pm }P(B_{i}=b|a)\mathrm{log}_{2}P(B_{i}=b|a).\eeq One can
again \cite{app} relate this to a bit error rate (assuming it is again equal
in both bases), which reduces to \beq 1-2h(R_{\mathrm{err}})\leq 0,\eeq where \beq h(R_{%
\mathrm{err}})=-R_{\mathrm{err}}\mathrm{log}_{2}R_{\mathrm{err}}-(1-R_{%
\mathrm{err}})\mathrm{log}_{2}(1-R_{\mathrm{err}}).\eeq

Ultimately, one may consider temporal steering inequalities as a benchmark
for validating the usability of a quantum channel for BB84. The connection
discussed here, between BB84 and steerability, is both natural and physical,
because of the Choi-Jamiolkowski isomorphism \cite{choi} and the known
symmetry between BB84 and entanglement based protocols \cite{Ekert}. However
it allows one to consider BB84 protocols \cite{app,1sided} in a practical
and direct way. Possible generalization to other protocols \cite{Bruss}
deserves further investigation.

Finally, one may ask what are the relative merits of steerability versus
full Bell non-locality in terms of their ability to characterize a quantum
channel. For cryptography schemes based on entangled pairs, encoding using
two mutually unbiased bases, and an eavesdropper strategy based on an
independent attack, both the Bell inequality violation, the steering
inequality violation, and BB84 are limited by the same bit error rate ($%
14.6\%$). In the entangled-based spatial scenario Alice and Bob typically
attempt to share an entangled singlet state. The effect of Eve inducing
errors on this state, due to measuring equally in two-bases, means that the
state that Alice and Bob receive is a Werner state. The Werner state is one
of the few states where the CHSH and steering inequality violations
coincide. The result by Branciard \emph{et al.} \cite{app} discussed above
suggests that in other situations steering inequalities, and thus steering
as a resource, may be more useful for validating quantum channels than Bell
non-locality (which is necessarily stricter).

\begin{figure}[th]
\includegraphics[width=7.5cm]{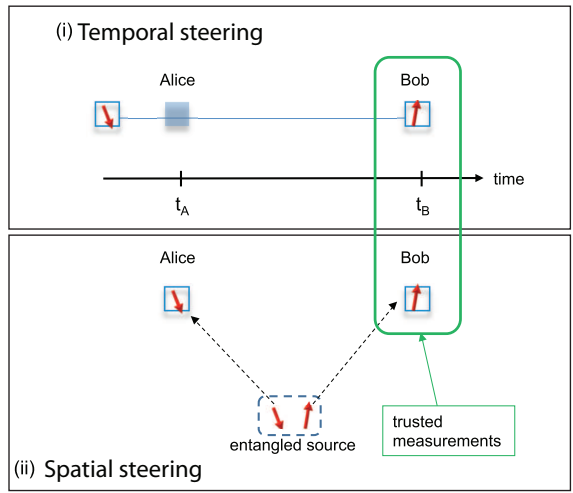}
\caption{(Color online) Two possible situations that can lead to the
violation of the steering inequality: (i) Bob's particle is measured by
Alice at a earlier time; (ii) Bob's particle is entangled with another one,
on which Alice performs measurement.}
\label{fig5}
\end{figure}

\section{Temporal Ordering}

In a real scenario there is a further symmetry between the temporal and
spatial steering inequalities. In both cases Bob trusts only his measurement
results and asks Alice to provide her measurement outcomes to him for
comparison. From Bob's point of view, a violation of the inequality may have
two possible origins (excluding non-local communication between his and
Alice's apparatus) (i) Bob's particle is measured by Alice at a earlier time
(temporal steering), or (ii) Bob's particle is entangled with another one,
on which Alice performs measurements (spatial steering).

To distinguish these two cases, the following steps could be made. (a) When
Bob receives the particle, he\emph{\ should not} perform his measurement
immediately and also ask Alice not to perform her measurement (though of
course she may already have done so, and on Bob's qubit in the temporal
scenario). (b) Bob should ask Alice to perform her measurement following his
orders, e.g., along the $\widehat{x}$, $\widehat{y}$, or $\widehat{z}$
direction. (c) After Alice reports her measurement results, Bob then
performs the corresponding measurement. If Alice has already pre-measured
Bob's qubit then, unless Bob chooses the basis she happens to have already
pre-measured in, she can only make a random guess. On the other hand, for
the case (ii), Alice can still measure her qubit and Bob's results can be
steered to give violations.

\section{Conclusion}

In summary, we have shown that there exists a temporal scenario of the
steering inequality for a single object. A strong connection to the bound on
the error rate of a quantum channel in the BB84 protocol is pointed out and
may have potential applications in quantum communication~\cite{Smith}.

\acknowledgments

This work is supported partially by the National Center for Theoretical
Sciences and National Science Council, Taiwan, grant numbers NSC
101-2628-M-006-003-MY3, NSC 100-2112-M-006-017, and NSC 102-2112-M-005 -009
-MY3. FN is partially supported by the RIKEN iTHES Project, MURI Center
for Dynamic Magneto-Optics, JSPS-RFBR Contract No. 12-02-92100, Grant-in-Aid
for Scientific Research (S), MEXT Kakenhi on Quantum Cybernetics, and the
JSPS via its FIRST Program.

\appendix

\section{Non-Markovian environment}

\begin{figure}[th]
\includegraphics[width=8cm]{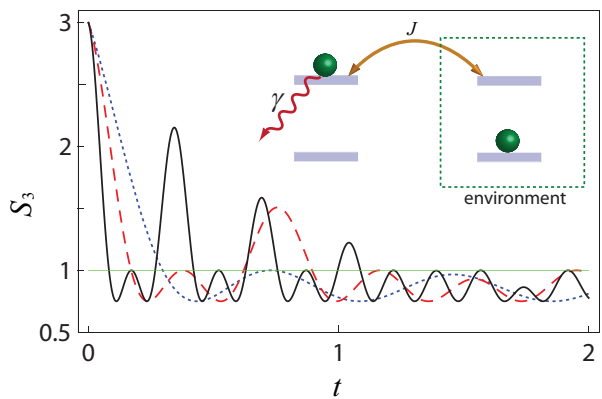}
\caption{(Color online) The temporal steering parameter $S_{3}$ as a
function of time. The black-solid, red-dashed, and blue-dotted curves
represent the results of $J=9\protect\gamma ,$ $4\protect\gamma $, and 2$%
\protect\gamma $, respectively. Inset: Schematic view of a qubit coupled to
a special environment (another qubit) with the coherent coupling strength $%
\hbar J$ and the intrinsic decay rate $\protect\gamma $. }
\label{fig3}
\end{figure}

When a system is coupled to an environment it is possible that coherence
between the system and the environment, i.e., entanglement, can be created
during the evolution. It is interesting to know if such a coherence can also
cause recurrent violations of the temporal steering inequality. To
investigate this we consider a single qubit coherently coupled to another
ancillary qubit, which serves as an effective environment, as shown in the
inset of Fig.~\ref{fig3}. If one traces out the effective-environment-qubit
the reduced system can be viewed as being coupled to a non-Markovian
environment.

In addition to this effective environment, we assume that the system is
subject to an intrinsic decay as in the example in the main text, but
without the influence of the Rabi-oscillation-inducing Hamiltonian. The
interaction between the system and the environment is written as $H_{\text{%
int}}=\hbar J(\sigma _{+}^{1}\sigma _{-}^{2}+\sigma _{+}^{2}\sigma _{-}^{1})$%
, where $\sigma _{+}^{i}$ and $\sigma _{-}^{i}$ are the raising and lowering
operators of the $i$-th qubit, and $\hbar J$ is the coherent coupling
between the system and the environment. The master equation of the total
system is
\begin{equation}
\dot{\rho}=\frac{1}{i\hbar }[H_{\text{int}},\rho ]+\frac{\gamma _{1}}{2}%
(2\sigma _{-}^{1}\rho \sigma _{+}^{1}-\sigma _{+}^{1}\sigma _{-}^{1}\rho
-\rho \sigma _{+}^{1}\sigma _{-}^{1}),
\end{equation}%
where $\gamma _{1\text{ }}$is the decay rate of the system qubit. In Fig.~%
\ref{fig3}, we plot the steering parameter $S_{3}$ for various coupling
strengths $\hbar J$. Similar to the previous example, Fig.~2 in the main
text, there are violations in the short-time regime. If the coupling $\hbar
J $ is strong enough, it can also induce recurrent violations at later
times. Again, the coherent coupling $\hbar J$ causes the recurrence of
coherence in the system qubit. These coherent off-diagonal terms then give
the violations of the inequality.

\section{Characterizing the dephasing of a quantum channel}

It is interesting to know if there are other practical applications of the
temporal steering inequality. To gain further insight, we analyze the
various contributions the steering parameter $S_{3}$ for the example system
discussed earlier, of a single qubit undergoing Rabi oscillations and decay
processes (Fig.~2 in the main text). Interestingly, we find the contribution
from the measurement in the $\widehat{x}$ direction is\emph{\ independent}
of the coherent Rabi frequency $g$ and takes the simple form%
\begin{equation}
E\left[ \left\langle \widehat{B}_{i}\right\rangle _{A_{i=\widehat{x}}}^{2}%
\right] =\exp (-2\gamma t),
\end{equation}%
while the contributions from the measurements in the $\widehat{z}$ and $%
\widehat{y}$ directions,
\begin{equation}
E\left[ \left\langle \widehat{B}_{i}\right\rangle _{A_{i=\widehat{y}}}^{2}%
\right]~~ \text{and}~~ E\left[ \left\langle \widehat{B}_{i}\right\rangle
_{A_{i=\widehat{z}}}^{2}\right],
\end{equation}
are functions of both $g $ and $\gamma $.


This means that the measurements in the $\widehat{x}$ direction are
independent of the coherent tunneling amplitude $g$, and allow us to extract
information about the dephasing rate $\gamma $ of the channel if we know in
advance the axis of the coherent tunneling.

Generally speaking, $E\left[ \left\langle \widehat{B}_{i}\right\rangle
_{A_{i=\widehat{x}}}^{2}\right] $, $E\left[ \left\langle \widehat{B}%
_{i}\right\rangle _{A_{i=\widehat{y}}}^{2}\right] $, and $E\left[
\left\langle \widehat{B}_{i}\right\rangle _{A_{i=\widehat{z}}}^{2}\right] $
are influenced by both coherent and incoherent properties of the channel.
Therefore, the value of the steering parameter can be used as an approximate
indicator of how good a channel is. For example, if the value of the
steering parameter $S_2$ is very close to $2$ after the qubit passes through
the channel, one can expect it does not lose much of its coherence, and
hence dephasing and losses are low. One can in principle check $S_2$ with
and verify the quality of the channel, rather than doing full process
tomography\cite{Nielsen}. How this scales to larger dimensional systems is
an interesting open problem.

\section{Method for assessing steering and error rates in the BB84 protocol}

We now show a method for calculating the average error rate $R_{\mathrm{err}%
} $ and the temporal steering parameter $S_{2}$ in the BB84 protocol.
Throughout this supplementary material, the temporal behavior of all
quantities will be calculated in the Schr\"{o}dinger picture.

Let us describe the BB84 protocol in terms of a quantum channel. A quantum
object is initially prepared as a density matrix $\rho_{0}$. Subsequently,
Alice performs her projective measurement on $\rho_{0}$, interacting a probe
(ancillary) qubit with the object and measuring the probe system
destructively. As a result, the object becomes a pure state $\rho_{\mathrm{A}%
,i,a} = \Pi_{i,a} $, with probability
\mbox{
\(
P^{\rm Q}(A_{i,t_{\rm A}}=a)
=
{\rm Tr} \, (\Pi_{i,a}\rho_{0})
\)}. The operator $\Pi_{i,a}$ is the projector onto the eigenvector of the $%
i $th component of the $2\times 2$ Pauli matrix ($i=\widehat{z},\widehat{x}$%
). When $i=\widehat{z}$, for example, we find that $\Pi_{\widehat{z},a} = |%
\widehat{z},a\rangle \langle \widehat{z},a| $, with $\sigma_{z}|\widehat{z}%
,a\rangle = a |\widehat{z},a\rangle $ and $a=\pm 1$. Between Alice and Bob,
Eve tries to eavesdrop the information sent. Eve's action can be described
by a linear map. For an input state $\rho$, one may write this linear map as
the Kraus representation\,\cite{Nielsen},
\begin{equation}
\rho\ \mapsto\ \mathcal{E}(\rho) = (1-p-q)\rho + q\,\pi_{\widehat{z}}(\rho)
+ p\,\pi_{\widehat{x}}(\rho),
\end{equation}
with
\begin{equation}
\pi_{i}(\rho) = \sum_{a} \Pi_{i,a} \, \rho \, \Pi_{i,a},\; 0\; \le \;
p\;,\;q\; \le 1,
\end{equation}
and \mbox{
\(
p+q \le 1
\)}. Therefore, Bob receives the density matrix
\mbox{
\(
\rho_{{\rm B},i,a} = \mathcal{E}(\rho_{{\rm A},i,a})
\)}. The process is assessed by the fidelity
\begin{eqnarray}
F(\rho_{\mathrm{A},i,a},\,\mathcal{E}) &=& \mathrm{Tr} \, ( \sqrt{\rho_{%
\mathrm{A},i,a}}\; \rho_{\mathrm{B},\mu}\, \sqrt{\rho_{\mathrm{A},i,a}} )
\notag \\
&=& (1-p-q)\; + q \;\langle i,a \,| \pi_{z}(\rho_{\mathrm{A},i,a})|\, i,a
\rangle  \notag \\
&& + p\,\langle i,a \,| \pi_{x}(\rho_{\mathrm{A},i,a})|\, i,a \rangle .
\label{eq:sim_BB84}
\end{eqnarray}
When this quantity is unity, the protocol completely works for a specific
input state $\rho_{\mathrm{A},i,a}$. The error rate $R_{\mathrm{err}}$ is
the mean value of $(1-F)$ over all possible input states.

The fidelity (\ref{eq:sim_BB84}) is calculated as follows. First, we focus
on the formula
\begin{equation}
|\langle\, i,a |\, j,b \rangle | = \delta_{i,j}\delta_{a,b} + \frac{1}{\sqrt{%
2}}(1 - \delta_{i,j}).
\end{equation}
The second term comes from the fact that $|\widehat{z},a\rangle$ and $|%
\widehat{x},a\rangle$ are the elements of the mutually-unbiased basis in a
two-level system\,\cite{Wootters;Fields:1989}. Using this formula, we obtain
\begin{equation}
\langle i,b \,| \pi_{j}(\rho_{i,a})|\, i,b \rangle = \delta_{i,j} \mathrm{Tr}%
\, (\Pi_{j,b}\, \rho_{i,a}) + \frac{1}{2}(1-\delta_{i,j}) .
\label{eq:expec_eves_action}
\end{equation}
Furthermore, we have $\mathrm{Tr}\, (\Pi_{i,a}\,\rho_{\mathrm{A},i,a})=1 $,
since $\rho_{\mathrm{A},i,a} = \Pi_{i,a} $. Thus,
\begin{equation}
F(\rho_{\mathrm{A},i,a},\mathcal{E}) = 1 - q\,\frac{ 1 - \delta_{i,\widehat{z%
}}}{2} - p\,\frac{1-\delta_{i,\widehat{x}} }{2}.
\end{equation}
It indicates that errors occur whenever Eve's measurement operators do not
commute with Alice's density matrix. Since $F(\rho_{\mathrm{A},i,a},\mathcal{%
E})$ does not depend on $a$ and
\mbox{
\(
\sum_{a}\,P^{\rm Q}(A_{i,t_{\rm A}}=a)=1
\)
}, the average error rate is
\begin{equation}
R_{\mathrm{eff}} = \frac{1}{2} \sum_{i,a} P^{\mathrm{Q}}(A_{i,t_{\mathrm{A}%
}}=a) (1 - F) = \frac{1}{4}(p+q).
\end{equation}

We turn to the expectation value of $B_{i}$ conditioned by Alice's result,
\begin{equation}
\langle B_{i,t_{\mathrm{B}}} \rangle_{A_{i,t_{\mathrm{A}}}=a} = \mathrm{Tr}%
\, ( \sigma_{i}\, \rho_{\mathrm{B},i,a} ).
\end{equation}
Using Eq.~(\ref{eq:expec_eves_action}), we find that
\begin{eqnarray}
\langle B_{i,t_{\mathrm{B}}} \rangle_{A_{i,t_{\mathrm{A}}}=a} &=& (1-p-q)
\mathrm{Tr}\, (\sigma_{i}\,\rho_{\mathrm{A},i,a})  \notag \\
&& + q \sum_{b=\pm 1}b\, \langle i,b \,| \pi_{\widehat{z}}(\rho_{\mathrm{A}%
,i,a}) |\, i,b \rangle  \notag \\
&& + p \sum_{b=\pm 1}b\, \langle i,b \,| \pi_{\widehat{x}}(\rho_{\mathrm{A}%
,i,a})|\, i,b \rangle  \notag \\
&=& a[(1-p-q) + \delta_{i,\widehat{z}}\,q + \delta_{i,\widehat{x}}\,p] .
\label{eq:cexpec_BB84}
\end{eqnarray}
In this derivation, we also use the fact
\mbox{
\(
\sigma_{i,a}\,\rho_{{\rm A},i,a}=a\,\rho_{{\rm A},i,a}
\)}. Since $a=\pm 1$, the conditional expectation squared does not depend on
$a$. Thus, the temporal steering parameter is
\begin{eqnarray}
S_{2} &=& \sum_{i=\widehat{z},\widehat{x}} \sum_{a=\pm 1} P^{\mathrm{Q}%
}(A_{i,t_{\mathrm{A}}}=a)\, [ \langle B_{i,t_{\mathrm{B}}} \rangle_{A_{i,t_{%
\mathrm{A}}}=a} ]^{2}  \notag \\
&=& (1-p)^{2} + (1-q)^{2}.
\end{eqnarray}

\end{document}